\documentclass[aps,pra,preprint,superscriptaddress,amsmath]{revtex4-1} 

\usepackage[colorlinks,citecolor=blue,linkcolor=blue,urlcolor=blue]{hyperref} 
\usepackage{graphicx}
\usepackage{bm}
\usepackage{setspace,lineno}
\modulolinenumbers[5]

\begin{document}

\title{Scanning tunneling spectroscopy reveals a silicon dangling bond charge state transition}

\author{Hatem Labidi}
\email{hatem.labidi@ualberta.ca}
\affiliation{Department of Physics, University of Alberta, Edmonton, Alberta, T6G 2J1, Canada}
\affiliation{National Institute for Nanotechnology, National Research Council of Canada, Edmonton, Alberta, T6G 2M9, Canada}

\author{Marco Taucer}
\affiliation{Department of Physics, University of Alberta, Edmonton, Alberta, T6G 2J1, Canada}
\affiliation{Quantum Silicon, Inc., Edmonton, Alberta, T6G 2M9, Canada}

\author{Mohammad Rashidi}
\affiliation{Department of Physics, University of Alberta, Edmonton, Alberta, T6G 2J1, Canada}
\affiliation{National Institute for Nanotechnology, National Research Council of Canada, Edmonton, Alberta, T6G 2M9, Canada}

\author{Mohammad Koleini}
\affiliation{Department of Physics, University of Alberta, Edmonton, Alberta, T6G 2J1, Canada}
\affiliation{National Institute for Nanotechnology, National Research Council of Canada, Edmonton, Alberta, T6G 2M9, Canada}

\author{Lucian Livadaru}
\affiliation{Quantum Silicon, Inc., Edmonton, Alberta, T6G 2M9, Canada}

\author{Jason Pitters}
\affiliation{National Institute for Nanotechnology, National Research Council of Canada, Edmonton, Alberta, T6G 2M9, Canada}

\author{Martin Cloutier}
\affiliation{National Institute for Nanotechnology, National Research Council of Canada, Edmonton, Alberta, T6G 2M9, Canada}

\author{Mark Salomons}
\affiliation{National Institute for Nanotechnology, National Research Council of Canada, Edmonton, Alberta, T6G 2M9, Canada}

\author{Robert A. Wolkow}
\affiliation{Department of Physics, University of Alberta, Edmonton, Alberta, T6G 2J1, Canada}
\affiliation{National Institute for Nanotechnology, National Research Council of Canada, Edmonton, Alberta, T6G 2M9, Canada}
\affiliation{Quantum Silicon, Inc., Edmonton, Alberta, T6G 2M9, Canada}

\begin{abstract}
We report the study of  single dangling bonds (DB) on the hydrogen terminated silicon (100) surface  using a low temperature scanning tunneling microscope (LT-STM).
By investigating samples prepared with different annealing temperatures, we establish the critical role of subsurface arsenic dopants on the DB electronic properties.
We show that when the near surface concentration of dopants is depleted as a result of $1250^{\circ}C$ flash anneals, a single DB exhibits a sharp conduction step in its I(V) spectroscopy that is not due to a density of states effect but rather corresponds to a DB charge state transition. The voltage position of this transition is perfectly correlated with bias dependent changes in STM images of the DB at different charge states.
Density functional theory (DFT) calculations further highlight the role of subsurface dopants on DB properties by showing the influence of the DB-dopant distance on the DB state. We discuss possible theoretical models of electronic transport through the DB that could account for our experimental observations.

\end{abstract}

\maketitle
\newpage
\begin{spacing}{1}
\textbf{\tableofcontents}
\end{spacing}

\section{Introduction}
Silicon dangling bonds on hydrogen terminated surfaces were studied for decades initially because of their impact on microelectronic devices. These point defects correspond to desorbed hydrogen atoms from the otherwise passivated silicon surface. They can occur spontaneously on the surface where their density depends on the passivation method. In metal oxide semiconductor (MOS) devices, hydrogen desorption is stimulated by hot electrons at the $Si/SiO_2$ interface causing
degradation of device performance \cite{Lyding.1996}. STM investigations of the technologically relevant Si(100):H surface demonstrated the possibility to use the STM tip to induce hydrogen desorption \cite{Lyding.1994,Shen.1995}. The study of the desorption yields at different experimental conditions revealed a substantial isotopic effect \cite{Lyding.1996,Foley.1998}. This led to replacing hydrogen by deuterium in the fabrication process of MOS devices which allowed for the improvement of their reliability and lifetime \cite{Lyding.1996,Hess.1998,Lyding.1998}.

Further STM tip induced hydrogen desorption studies showed the optimization of this atom scale lithographic technique for the precise creation of single DB structures \cite{Soukiassian.2003,Tong.2006} as well as complex DB patterns \cite{Hitosugi.1998,Walsh.2009,Soukiassian.2003-2}.
This led to many studies which established the relevance of DBs as building block for various prospective nano-scale electronic devices. For example, it was shown that a DB can be considered as a single atom switch that can be activated by STM tip induced electronic excitation \cite{Quaade.1998,Bellec.2010}. Single DBs can also serve as specific adsorption sites for functionalized molecules \cite{Mayne.2004} or induce the self directed growth of molecular structures \cite{Lopinski.2000} and regulate the conductivity through these structures \cite{Piva.2005}.
Moreover, several theoretical studies explored the possibility for DB wires created with the STM tip to serve as interconnects for atomic or molecular devices \cite{Kawai.2012,Kepenekian.2013,Ample.2013}. 

Interestingly, other studies showed that a DB can be considered as a single atom quantum dot \cite{Haider.2009,Livadaru.2010,Schofield.2013,Taucer.2014} and thus can be used to build charge quantum bits (qubits) \cite{Livadaru.2010}, artificial molecules \cite{Schofield.2013,Wolkow.2014} quantum dot cellular
automata (QCA) cells \cite{Haider.2009,Pitters.2011-2} and QCA based circuits \cite{Haider.2009,Wolkow.2014}. The important property allowing such consideration is the DB's quantized charge state levels lying in the silicon band gap. In fact, a DB can be either empty, singly or doubly occupied which correspond respectively to the DB being positive($DB^+$), neutral ($DB^0$) or negative($DB^-$) \cite{Haider.2009,Livadaru.2011,Schofield.2013,Taucer.2014}. This makes conduction through the DB predominantly governed by tunneling rates through its charge states  \cite{Livadaru.2011,Schofield.2013,Nguyen.2010,Berthe.2008}, hence the quantum dot analogy. 

\begin{figure*}[ht]
	\centering
	\includegraphics[width=0.6\textwidth]{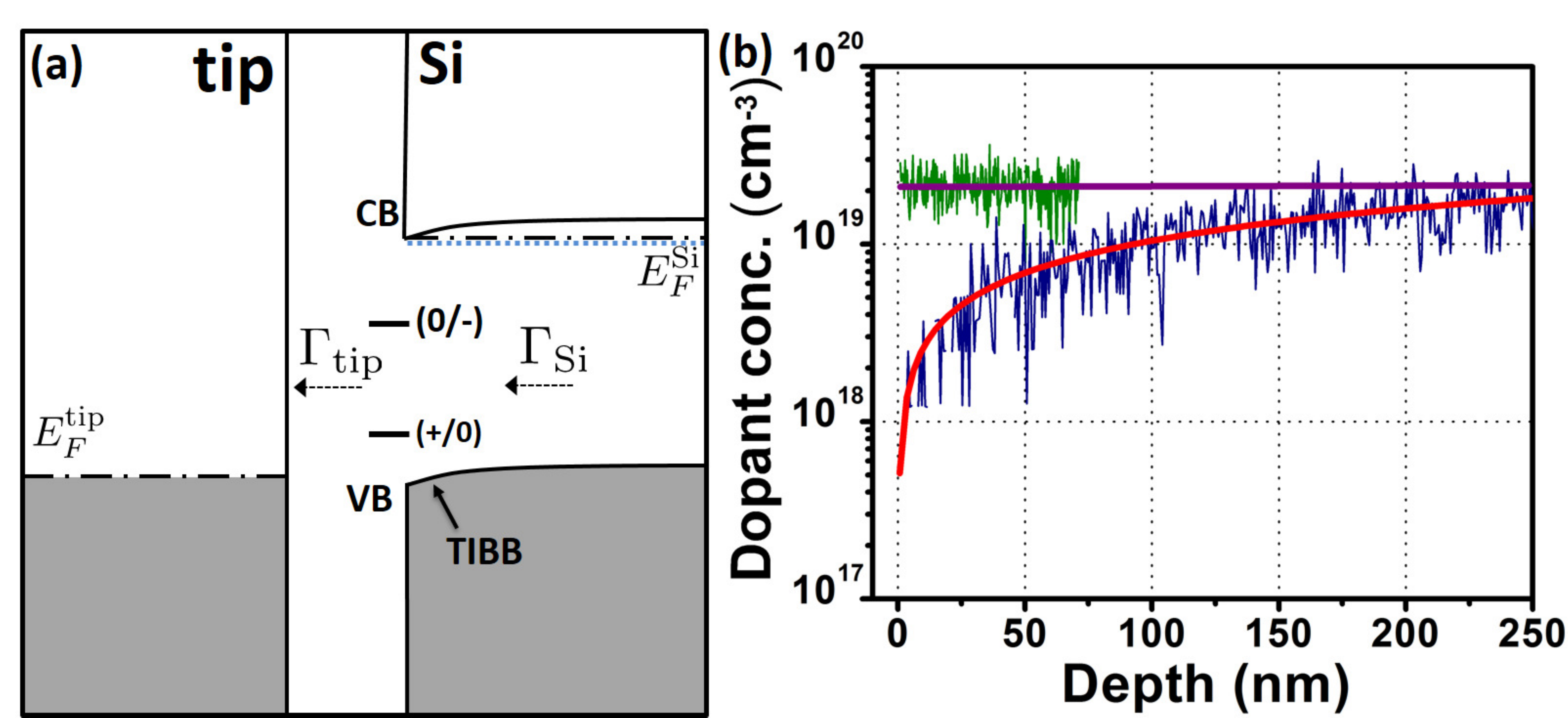}
	\caption{(a) Schematic energy band diagram of the STM junction showing the tunneling mechanism through a single DB. (+/0), (0/-) are respectively the charge transition levels of the DB from positive to neutral and from neutral to negative. $\Gamma_{tip}$ and $\Gamma_{Si}$ denotes the tunneling rates from sample to DB and from DB to tip respectively. The blue dotted line indicates the donor band and the dot-dashed lines indicate the Fermi levels. The tip induced band bending (TIBB) is indicated. Additional band bending effects, induced by the charge state of the DB, may also occur, but these are omitted from band diagrams throughout this manuscript. (b) Semi-logarithmic SIMS depth profiles of arsenic atoms for samples prepared with $1050^{\circ}C$ (green) and $1250^{\circ}C$ (blue) flashes; purple and red curves are the corresponding fits. Both samples, with an initial resistivity of $3-4\, m\Omega\, cm$, were flashed 2 times for about 1 second for each flash.}
	\label{Fig1}
\end{figure*}

In figure \ref{Fig1}-a we illustrate the general tunneling mechanism through a single DB. We define the energy levels (+/0) and (0/-) as the charge transition levels of the DB from positive to neutral and from neutral to negative respectively. When probing filled states, $\Gamma_{Si}$ represents the tunneling rate from sample to DB and $\Gamma_{tip}$ that from DB to tip. The STM tip acts as a gate by locally bending the silicon bands; an important effect known as tip induced band bending (TIBB) that was extensively studied in the case of semiconductor surfaces \cite{Feenstra.2003,Livadaru.2011,Teichmann.2008,McEllistrem.1993,Feenstra.2007} and was shown to play a major role in conduction through the DB in empty states \cite{Bellec.2010,Livadaru.2011,Pitters.2012,Schofield.2013,Piva.2014}.
In this configuration, $\Gamma_{tip}$ strongly depends on the tip-sample distance, i.e. tunnel current. On the other hand, $\Gamma_{Si}$ depends predominantly on the silicon dopant concentration but also on the tunnel junction voltage and the TIBB.

On highly doped samples, the dopant concentration near the surface can be ``tuned" depending on the particular thermal treatment the samples undergoes during cleaning. This is clearly seen from secondary ion mass spectrometry (SIMS) experiments shown in figure \ref{Fig1}-b.  As discussed in detail recently by Pitters \textit{et al.} \cite{Pitters.2012}, flashes at $1250^{\circ}C$ induce a dopant depletion region with the subsurface arsenic atom concentration dropping by almost 2 orders of magnitudes in the shallow subsurface regions. On the other hand, flashes at $1050^{\circ}C$ do not induce such dopant concentration changes. At room temperature, this was already demonstrated to have an impact on the DB conduction in empty states imaging \cite{Pitters.2012,Piva.2014}.

In this article, we study the influence of subsurface  arsenic dopants on the electronic properties of single DBs at low temperature. 
We show that depending on the thermal treatment of the sample, the DBs exhibit different STM imaging and spectroscopy characteristics in filled states. We report for the first time the direct observation of a DB charge state transition in the case of samples with a substantial subsurface dopant depletion layer. We observe a conduction step in I(V) scanning tunneling spectroscopy (STS) of single DBs which translates into a sharp peak in dI/dV. We show that the voltage position of this peak is perfectly correlated with bias dependent STM imaging of the DB that exhibits three distinct regimes:
(i) a rate-limited regime for voltages lower than the onset of the peak, when the DB is predominantly positively charged, exhibits a dark-halo in topography and does not carry current; and (ii) a transition regime for voltages at the peak onset when the bulk to DB current increases, the DB toggles between positive and negative states, starts to carry current and appears striated in topography; then (iii) for voltages past the peak onset, the DB stays negatively charged and appears as a bright protrusion in topography. We find that the voltage threshold of the charge state transition, i.e. dI/dV peak, can be different from one DB to another even on the same sample. Through DFT calculations, we show that this could be explained by the influence of the arsenic subsurface dopants on the DB state. We discuss possible theoretical models of electronic transport through the DB that could explain the charge state transition mechanism.

\section{Experimental methods}
Experiments were carried out using a customized commercial LT-STM (Omicron) operating at $4.5\, K$.
We used tungsten tips DC etched from a $0.25\, mm$ polycrystalline wire. Tips were cleaned in ultra-high vacuum (UHV) by a series of electron beam and field emission followed by field evaporation and sharpening in a field ion microscope (FIM), ensuring very small radius of curvature and robust tips \cite{Rezeq.2006,Pitters.2013}. 

Samples were cleaved from highly arsenic doped silicon Si(100) wafers (Virginia Semiconductor Inc.) with a resistivity of $3-4 \, m\Omega\, cm$ ($\sim 1.5\times 10^{-19}\, atom\, cm^{-3}$). Upon loading in a small UHV chamber with a base pressure of  $3\times10^{-11}\,Torr$, the sample was set to degas for about 12 hours at $\sim 600^{\circ}C$. Following a first rapid flash anneal at $\sim 900^{\circ}C$ to remove the oxide layer, high temperature flashes at $1050^{\circ}C$ or $1250^{\circ}C$ were applied for about 2 seconds, then the resistive heating is quickly shut down. The sample was set to cool between each flash until the base pressure is recovered.
After about 4 flashes, pure molecular hydrogen gas ($H_2$) was leaked into the UHV chamber. A hot tungsten filament ($\sim 1800^{\circ}C$) facing the sample was used to crack molecular hydrogen. We first leave atomic hydrogen to etch the silicon surface \cite{Wei.1995} for about 2 minutes prior to setting the sample temperature at $\sim 330^{\circ}C$ for an additional 2 minutes to ensure a $Si(100):H-2\times 1$ reconstruction \cite{Boland.1993,Mayne.2006,Bellec.2009,Pitters.2012}. Shortly after turning off the filament, the resistive heating and closing the leak valve, the sample is transferred to the LT-STM scanner.
STM images of large areas show that this preparation method generally yields a high quality surface with very low defect density and large terraces. 

DBs could be either found natively on the surface or created with the STM tip \cite{Shen.1995,Foley.1998}. To create a single DB, a small area is imaged then the tip is placed on top of a hydrogen atom and the feedback loop is switched off. A voltage pulse of $2\, ms$ at $+2.3\,V$ or higher (depending on the tip) is then applied. A sudden increase in the tunnel current during the pulse usually indicates the creation of the DB which can be confirmed by a subsequent topography of the same area \cite{Soukiassian.2003,Tong.2006,Haider.2009,Bellec.2010,Schofield.2013,Kolmer.2014}.

The different I(V) spectroscopies presented in this paper were acquired with the feed-back loop held off, i.e. at a constant tip-sample distance. An internal lock-in (Nanonis) was used to acquire differential conductance spectroscopy (dI/dV), with a modulation signal of $50\, mV$ amplitude and $875\, Hz$ frequency. All experiments were done with the sample grounded and the bias applied to the tip. Spectra were recorded for a voltage range of [-2.2;+1.8]V to avoid inducing the switching of the DB from one side to the other of a silicon dimer\cite{Quaade.1998,Bellec.2010} or hydrogen desorption \cite{Haider.2009,Bellec.2010} during data acquisition. Different in-situ tip improvement techniques such as high voltage pulses and gentle controlled crashes were used until a high quality tip was obtained for both spectroscopy and STM imaging in filled and empty states. Spectroscopy on bare silicon areas and bare dimers were used as a reference to ensure the absence of artifacts in spectroscopy due to the tip structure \cite{Dubois.2005,Bellec.2009}.  

\section{Minimal subsurface dopant depletion: \boldmath{ $1050^{\circ}C$} flashed samples }
\begin{figure*}[ht]
	\centering
	\includegraphics[width=0.6\textwidth]{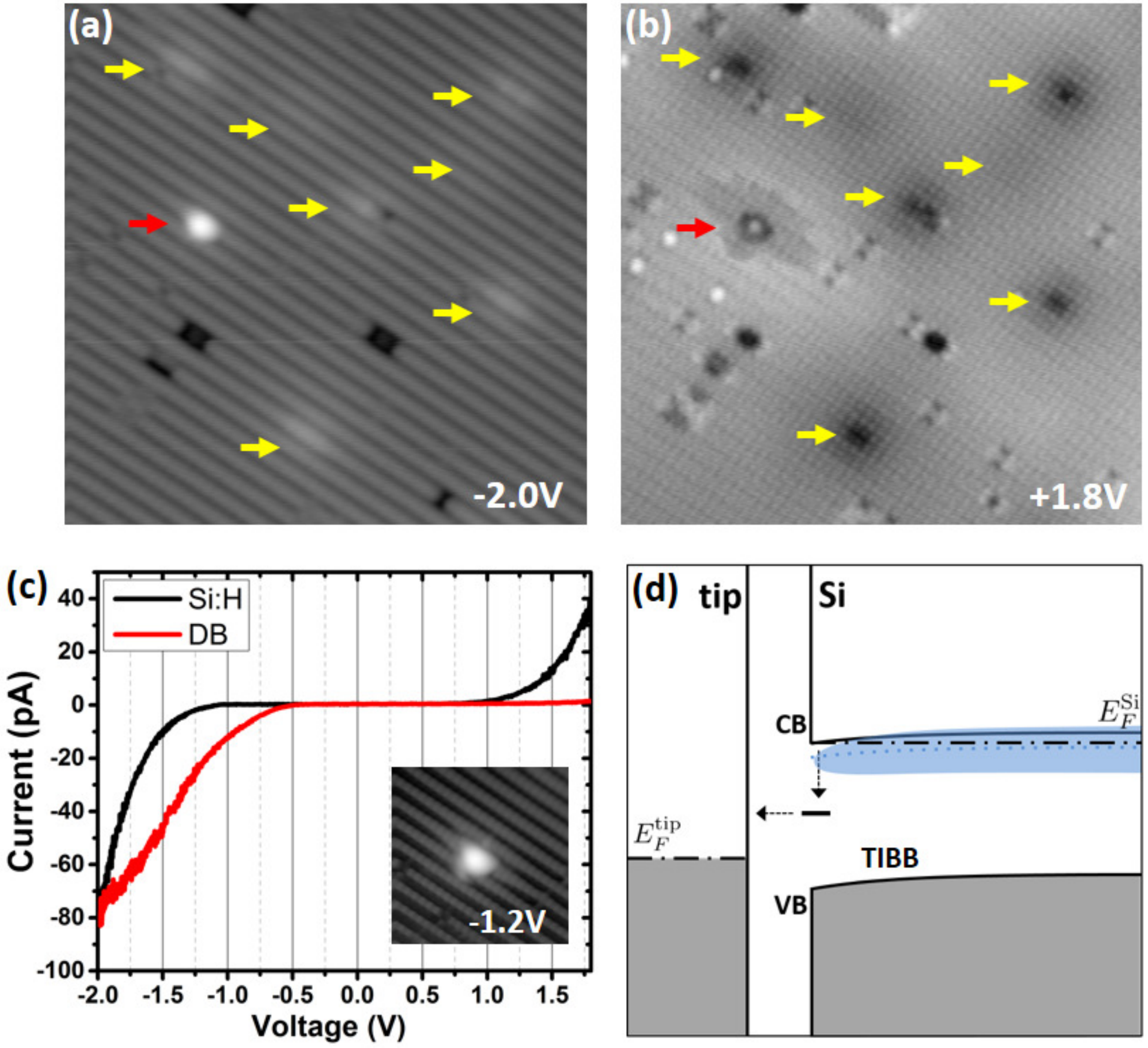}
	\caption{$(20\times20)nm^2$ constant current ($30\,pA$) filled state (a) and empty state (b) STM images of a $Si(100):H-2\times1$ surface of a $1050^{\circ}C$ flashed sample. Yellow arrows indicate subsurface dopants (As) while the red arrow points to a single DB. (c) I(V) spectroscopy acquired on the single DB (red curve) and the Si:H surface (black curve). The inset on (c) shows a constant current STM image of a single DB appearing bright at $-1.2\,V$. (d) Schematic energy band diagram showing tunneling mechanism through the DB in case of a $1050^{\circ}C$ flashed highly doped sample. Dash-dotted lines represent the Fermi levels of the tip and sample, $E_{F}^{tip,Si}$. The shaded blue region indicates the formation of a donor band which merges with the CB. }
	\label{Fig2}
\end{figure*}

Figure \ref{Fig2}-a and b show typical filled and empty state images of a surface prepared with flashes at $1050^{\circ}C$.  We notice the high concentration of arsenic subsurface dopants (indicated by yellow arrows) recognizable from their particular topography in both filled and empty states. The observation of subsurface dopants in STM images was already reported in several articles \cite{Bellec.2008,Piva.2014} and their electronic properties recently studied in detail by Sinthiptharakoon \textit{et al.} \cite{Sinth.2014}. Their near surface concentration can be estimated directly from STM images as explained by Piva \textit{et al.} \cite{Piva.2014}.  For the sample studied in figure \ref{Fig2}, we find $\sim 1\times 10^{-19}\,atom\,cm^{-3}$ which is consistent with the  SIMS data in figure \ref{Fig1}-b and previous studies that showed no significant subsurface dopant depletion for samples flashed at $1050^{\circ}C$ \cite{Pitters.2012,Sinth.2014,Piva.2014}.

A single DB (indicated by a red arrow in Figure \ref{Fig2}-a and b) appears as a bright protrusion in filled states and exhibits the characteristic dark halo surrounding it in empty states \cite{Liu.2002,Haider.2009,Bellec.2009,Livadaru.2011,Pitters.2011,Schofield.2013,Piva.2014}.  
Figure \ref{Fig2}-c shows comparison of typical I(V) spectroscopies acquired on a single DB and the Si:H surface on the $1050^{\circ}C$ flashed samples. We can notice that the DB starts to carry current at voltages before the STM measured valance band (VB) maximum at around $-1.15\,V$. Furthermore, images at different voltages reveal that DBs appear always bright even at voltages as low as $-1.2\,V$ (close to VB edge) as seen in inset image on figure\ref{Fig2}-c. These STM and STS data are similar to what was reported in previous studies of DBs on highly doped samples at room temperature \cite{Liu.2002,Piva.2014}.

 In degenerately doped n-type silicon ($N_d \gtrsim10^{19}\, cm^{-3}$) at cryogenic temperatures, as with n-type medium- or highly-doped silicon at room temperature, the conduction band is populated by a substantial density of itinerant electrons. At room temperature, these are thermally promoted to the conduction band and distributed roughly according to a Boltzmann distribution. At cryogenic temperatures, however, itinerant electrons occupy the disordered donor band which, when doping is degenerate, merges with the conduction band and allows electrons to delocalize across the crystal \cite{shklovskii1984}. Thus, for room temperature studies, and for low temperature studies of samples flashed to only $1050^{\circ}C$, the itinerant electrons exist throughout the entire crystal and extend all the way to the surface. 

Therefore, and as illustrated in figure \ref{Fig2}-d, surface DBs are supplied by a large number of electrons from the conduction band (CB) with sufficient energy to populate the DB and keep it negative. In this case, the emptying rate $\Gamma_{tip}$ cannot overcome the filling rate from the sample $\Gamma_{Si}$ leaving the DB negative during normal imaging and STS conditions. In fact, on the $1050^{\circ}C$ flashed samples, we saw that almost all the single DBs we studied start to carry current at voltages as low as $-0.5\,V$ in the example of figure 2-c, well below the VB onset ($-1.15\,V$). At such low voltages, no surface nor silicon states exist, aside from the DB itself, which strongly suggests that the DB is being supplied with electrons from the CB.

\section{Large subsurface dopant depletion: \boldmath{$1250^{\circ}C$} flashed samples}
\begin{figure*}[ht]
	\centering
	\includegraphics[width=0.6\textwidth]{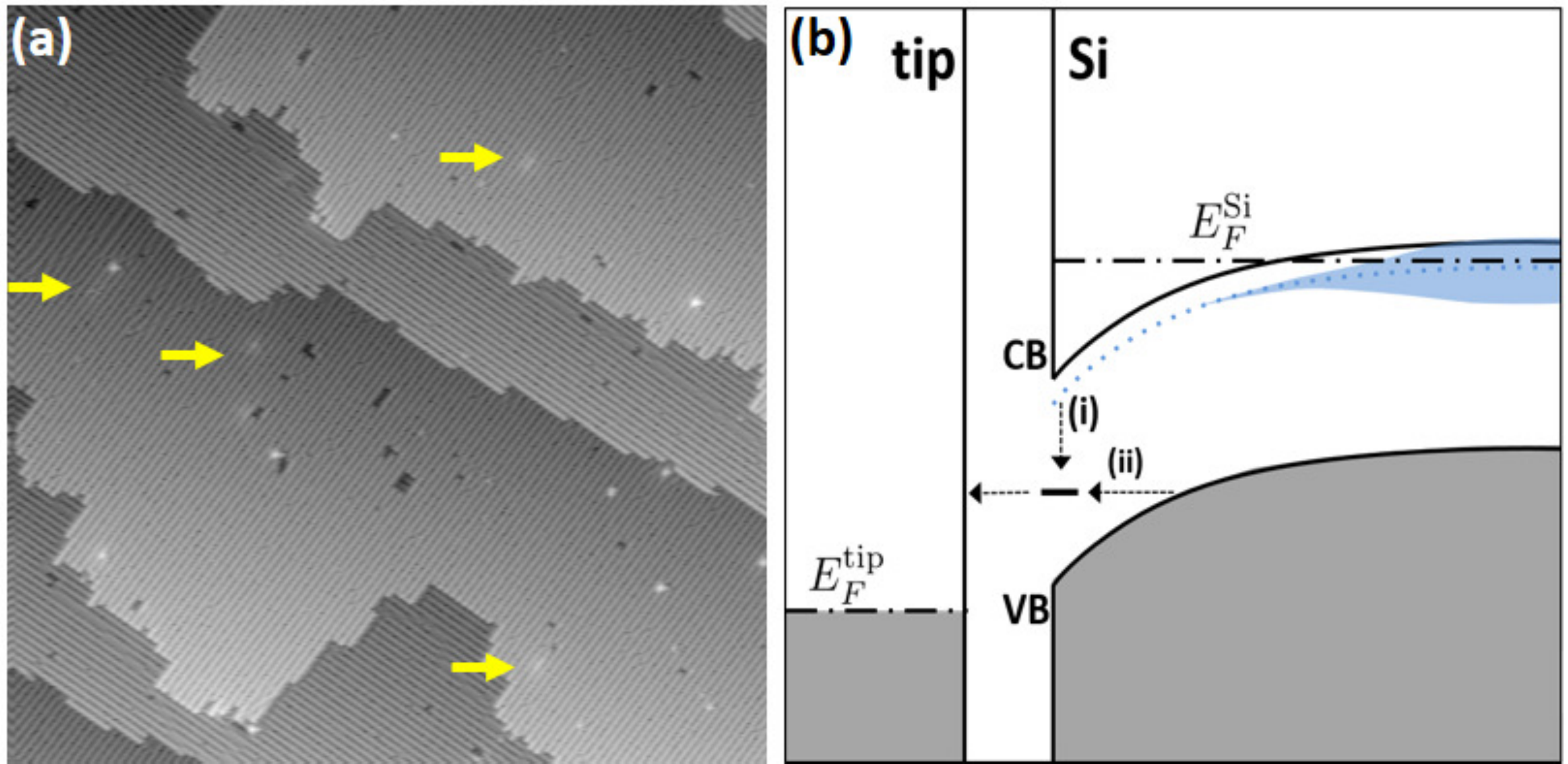}
	\caption{(a) $(80\times 80)nm^2$ constant current ($40\,pA$) filled state STM image at $-2.0\,V$. Yellow arrows indicate subsurface dopants. (b)  Schematic energy band diagram showing possible dynamics during STS of a DB in occupied states for a sample flashed to $1250^{\circ}C$. Dash-dotted lines represent the Fermi levels of the tip and sample, $E_{F}^{tip,Si}$, and the blue dotted line represents the energy of the bound state of a donor. The shaded blue region indicates the formation of a donor band which merges with the CB. Roman numerals represent processes which extract or inject electrons from the DB. Energies and TIBB are exaggerated for clarity.}
	\label{Fig3}
\end{figure*}

Figure \ref{Fig3}-a shows a typical filled states STM image of a relatively large surface area of a sample prepared with $1250^{\circ}C$ flashes. We clearly notice that the concentration of subsurface dopants, estimated at $\sim 8\times 10^{-17}\,atom\,cm^{-3}$ in this example, is considerably reduced compared to the $1050^{\circ}C$ flashed sample which is consistent with SIMS data (figure \ref{Fig1}-b) showing a subsurface dopant depletion region. In this case, instead of forming a donor band, donors in the subsurface dopant depletion region have a localized hydrogen-like bound state. This is illustrated in the schematic energy band diagram in figure \ref{Fig3}-b.
In the following, we examine carefully the impact of this subsurface dopant depletion on the electronic properties of single DBs.

\subsection{Influence on STM imaging of single DBs}
In unoccupied states images, similarly to the previous case of $1050^{\circ}C$ flashed sample, a single DB still appears as a slightly bright protrusion surrounded by a dark halo as seen in figure \ref{Fig4}-a. This characteristic image of a single DB on the Si:H surface for voltages below $+2\,V$ \cite{Liu.2002} is due to the negative charge of the DB. In fact, this localized negative charge distorts locally the electronic bands of the silicon sample, which in turn affects the STM current in that vicinity causing the halo effect. This was modeled in an earlier study by Livadaru \textit{et al.} \cite{Livadaru.2011} and was shown to arise from the nonequilibrium charging of the DB during STM imaging in unoccupied states. The dynamic aspect of it could be accessed experimentally using an LT-STM by studying the telegraph noise seen on the edge of the halo surrounding the DB \cite{Taucer.2014}.

In filled states images, the same DB appears as a bright protrusion  at $-1.8\,V$ (Figure \ref{Fig4}-b). This was believed to be the case regardless of the imaging voltage and was explained by the fact that a single DB in equilibrium with the sample is always negatively charged in filled states imaging of highly doped n-type Si:H \cite{Liu.2002,Bellec.2009,Schofield.2013}, similarly to what we saw previously in $1050^{\circ}C$ flashed samples. 
However, we clearly see from figure \ref{Fig4}-c that when we decrease the bias, the DB becomes dark with a small halo surrounding it; a signature of an electrostatic effect due to a localized charge and strongly suggests that the DB is positive \cite{Brown.2003,Livadaru.2011,Schofield.2013}. Indeed, since we are probing occupied states (negative voltage on sample), a localized positive charge on the surface will induce a local downward bending of the silicon bands. This influences the electronic conduction through the DB and induces a rate limited tunneling regime\cite{Livadaru.2011} resulting in a much lower current when the STM tip scans over the DB. 

\subsection{Scanning tunneling spectroscopy: identification of charge state transition}
\begin{figure*}[ht]
	\centering
	\includegraphics[width=0.8\textwidth]{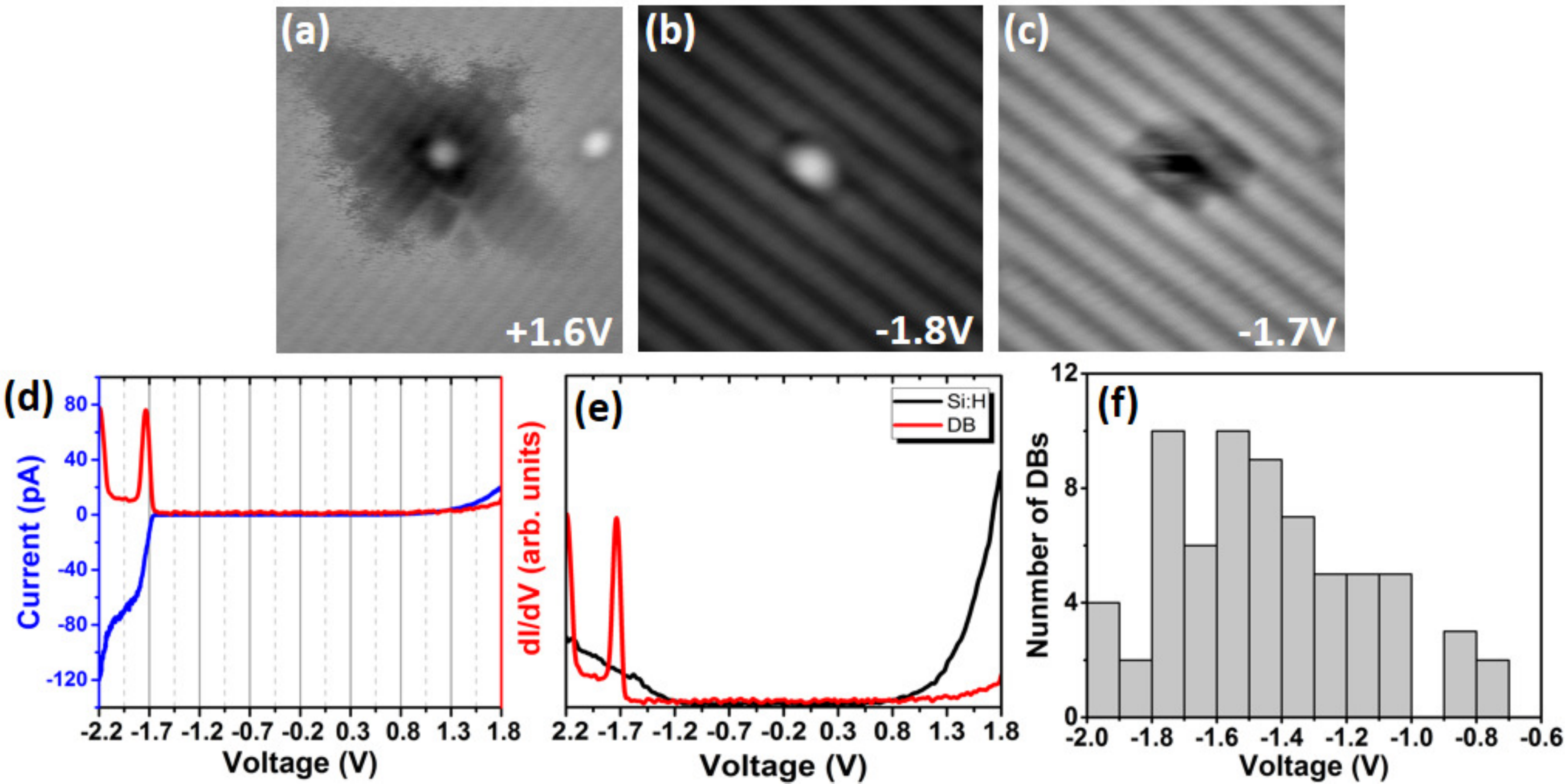}
	\caption{(a), (b) and (c): $(10 \times 10)nm^2$ constant current ($30\,pA$) STM images of a single DB at $+1.6\,V$, $-1.8\,V$ and $-1.7\,V$, respectively. (d) I/V (blue curve) and dI/dV (red curve) spectroscopy of the same DB. (e) dI/dV spectroscopies of a single DB (red curve) and Si:H surface (black curve). (f) Statistic over 69 DBs showing the variation of the the charge state transition peak voltage onset.}
	\label{Fig4}
\end{figure*}

In order to gain more information on the possible origin of this voltage dependent imaging, we performed STS measurements using the parameters and procedures described in the experimental section. Figure \ref{Fig4}-d shows I(V) (blue curve) and dI/dV (red curve) spectroscopy acquired on the same DB imaged in figure \ref{Fig4}-a. The I(V) curve shows a conduction step that translates into a very sharp peak in the dI/dV spectroscopy with less than $0.1V$ width. Figure \ref{Fig4}-e allows direct comparison of the dI/dV spectroscopy acquired with the STM tip on top of the single DB (red curve) and on top of the Si:H surface (blue curve). We can clearly see that the sharp peak appears only on the DB spectra and not elsewhere on the passivated surface. The dI/dV spectra measured on Si:H surface is also consistent with what was reported in previous studies in the literature \cite{Bellec.2009,Liu.2002}. However, unlike those studies, we see in our case that the apparent surface band gap measured on the single DB spectra seems to be larger than what is measured on the surface.  

The striking result when comparing STS and STM images in figure \ref{Fig4} is the position of the sharp transition peak located at $-1.75\,V$ in this example, as it becomes obvious that the STS is consistent with the voltage dependent imaging of the single DB. Indeed, for voltages larger than the onset of the peak maximum, the DB appears as a bright protrusion (figure \ref{Fig4}-b) but for lower voltages (closer to $0\,V$) the DB appears dark and surrounded with a halo (figure \ref{Fig4}-c). This trend was consistently confirmed in all the studied DBs and through different samples prepared with  $1250^{\circ}C$ flashes as described in the experimental methods section. 
Therefore, the sharp STS peak does not correspond to a density of state feature but rather to the voltage onset for the charge state transition of the DB, which is similar to the characteristic charge transition peaks seen in I(V) curves of a single electron transistor.

Even though previous studies already discussed STS of single DBs on n-type Si(100):H \cite{Liu.2002,Bellec.2009,Ye.2013,Kolmer.2014}, the STS features presented in this article and the related voltage dependent evolution of STM images in filled states were not previously reported in the literature. Bearing in mind the previous results obtained on $1050^{\circ}C$ flashed samples, it becomes clear that the key point for our observations is the subsurface dopant depletion layer induced by the sample thermal treatment. In fact, since the degenerate region does not reach the surface as illustrated in the band diagram of figure \ref{Fig3}-b, DBs are poorly supplied with electrons from the sample. Therefore, the DB can be more closely coupled to electronic states of the tip, where it sees resonant energy levels, than with the sample, where it is isolated in the mid-gap. In such cases, the occupation of the DB energy level is not necessarily determined by the sample Fermi level. Instead, the occupation of the DB level is determined by a competition of rates, as electrons are supplied from the sample ($\Gamma_{Si}$), and extracted by the tip ($\Gamma_{tip}$). A discussion on the electronic transport throught the DB and the charge transition mechanism is presented in the last section of this paper. 

It is important to note here that through our experimental investigation of hundreds of DBs on different samples, we saw that the charge state transition voltage onset can be different from one DB to another. Clearly, the sample preparation conditions, in particular flashing time and temperature, play the most important role. However, DBs' charge state transition onset varies even on a same sample. Figure \ref{Fig4}-f shows a statistic over 69 DBs on 2 samples prepared in the same way. We clearly see that the onset voltage varies from -0.8V to -2.2V. This variation can be explained by the non-uniform distribution of subsurface dopants which can influence the DB state and the electronic conduction through the DB. Other factors such as local surface defects may also play a role.
 
We also note that some DBs exhibit other STS features besides the sharp charge state transition peak, e.g. second peak in figure \ref{Fig4}-d. These features are different in nature from the first peak as they do not correspond to the previously described charge state transition correlated with STM images. Moreover, we noticed that DBs can exhibit multiple additional peaks or none at all. We think that these additional STS features could correspond to bulk impurity or surface states that become accessible as a consequence of an increased tunnel junction voltage and they depend therefore on the DB local environment.

\subsection{Evolution of DB imaging with voltage in filled states}
\begin{figure*}[ht]
	\centering
	\includegraphics[width=0.8\textwidth]{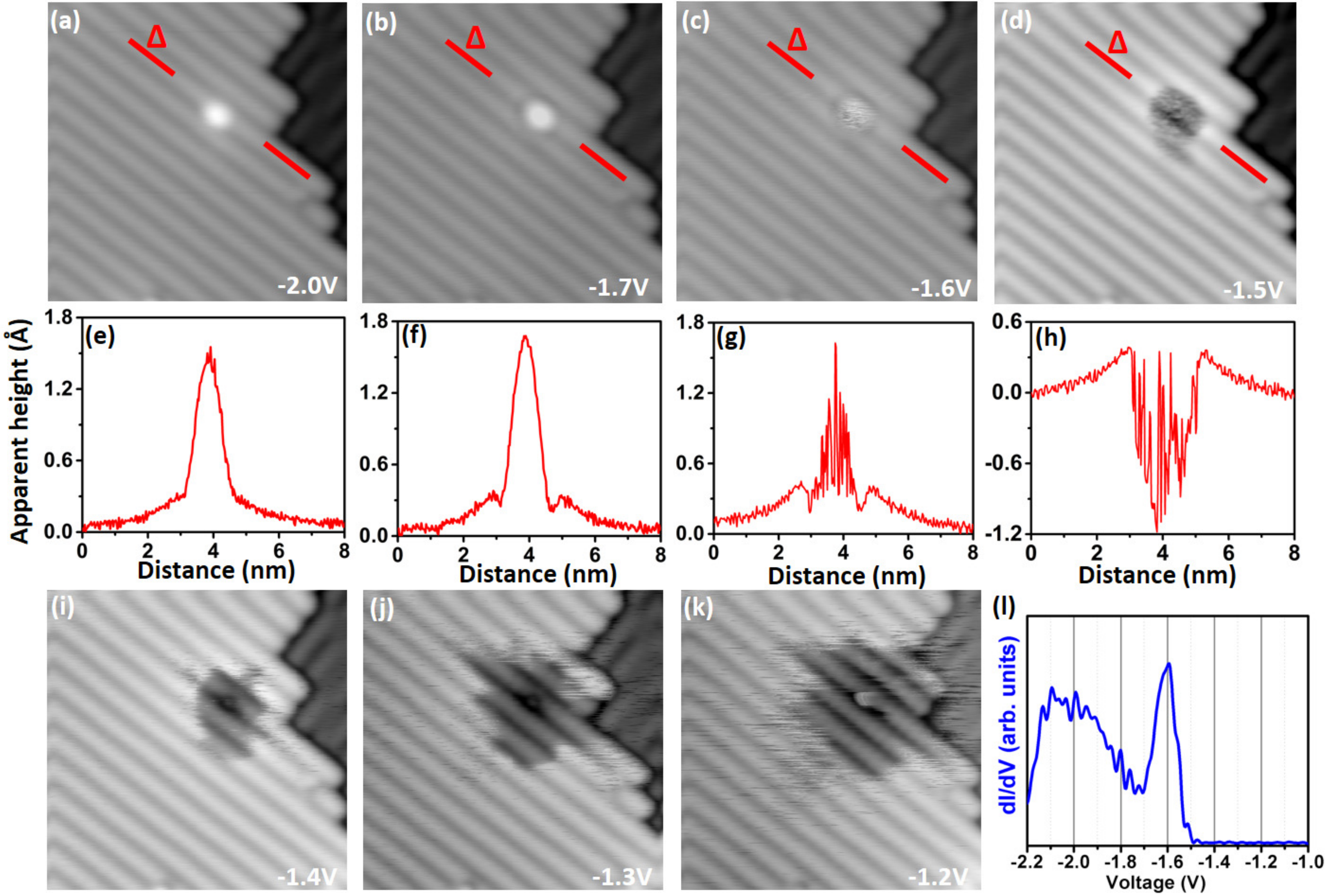}
	\caption{Series of $(10 \times 10)nm^2$ STM images of a single DB in filled 
		states at (a) $-2.0\,V$, (b) $-1.7\,V$, (c) $-1.6\,V$ and (d) $-1.5\,V$. (e) to (h): apparent height profile measured through the axis $\Delta$ of the corresponding above STM image. (i) to (k): STM images ($(10 \times 10)nm^2$ and $30\,pA$) of the same DB showing extended halo for higher voltages. (l) dI/dV spectroscopy measured on this specific DB}
	\label{Fig5}
\end{figure*}

Figure \ref{Fig5} shows a series of filled state STM images of a single DB at different voltages as well as the corresponding line profiles across the DB through the axis $\Delta$.
At $-2.0\,V$ (figure \ref{Fig5}-a), the DB appears as a localized and bright protrusion. The corresponding line profile (figure \ref{Fig5}-e) shows an exponential decay of the apparent height with radial distance from the DB, consistent with a localized negative charge state \cite{Hamers.1988,Schofield.2013}.
When decreasing the scanning voltage to $-1.7\,V$ (figure \ref{Fig5}-b), we notice a more spatially extended imaging of the DB with the appearance of a small halo surrounding it. This is more clear in the corresponding line profile (figure \ref{Fig5}-f) where we see a small decrease in the apparent height before the maximum located at the center of the DB.

In figure \ref{Fig5}-c, the voltage was decreased to $-1.6\,V$ and the change in the image is more dramatic: the DB no longer appears bright as in higher voltages but rather striated. The corresponding line profile (figure \ref{Fig5}-g) shows a noisy apparent height that toggles between a maximum corresponding to the same maximum height recorded for higher voltages, 1.6\AA{} in this example, and a certain minimum around 0.4\AA{}. When looking at the dI/dV spectra acquired for this DB (figure \ref{Fig5}-l), we see that the onset of the sharp peak ascribed the DB charge state transition is located exactly at $-1.6\,V$. Therefore, the striated imaging of the DB is consistent with a change in the conductivity during scanning as the DB charge state toggles between positive and negative.

At $-1.5\,V$ (figure \ref{Fig5}-d), which corresponds to the tail of the charge transition peak, we see that the DB image becomes more extended and dark but still striated. This is further highlighted in the corresponding line profile (figure \ref{Fig5}-h) where we see the profile of a halo similar to what was reported for the DB empty states imaging. Additionally, we also see that the apparent height inside the halo toggles between a minimum and a maximum which is consistent with the striated STM images. In fact, since the voltage is still at the tail of the peak, it's reasonable to assume that due to the tunneling rate competition the DB still undergoes an intermittent charge state transition.

If we further reduce the voltage to $-1.4\,V$, $-1.3\,V$ and $-1.2\,V$ as seen in figure \ref{Fig5}-i,j and k respectively, the halo surrounding the dark DB becomes more and more extended. This can be explained by the effect of the TIBB similarly to what was reported for empty states imaging of DBs \cite{Schofield.2013,Taucer.2014}. Indeed, the TIBB and the tunneling current from the silicon sample increase as the voltage is increased.  As a consequence, the tip must get closer to the DB so that $\Gamma_{tip}$ becomes larger than $\Gamma_{Si}$, resulting in a smaller dark halo. 

We can notice here that even though the current in STS appears to be zero, we can still image the DB. Actually, between the valance band maximum (VBM) and the charge state transition voltages, the current is not strictly zero. In fact, when we acquire STS, the feedback loop is switched off at -2.0V where the DB appears bright. When sweeping down the voltage while the tip-sample distance is kept constant, the DB eventually becomes positive and does not carry as much current as its negative state. Therefore, the current before the STS peak appears to be zero in  STS but is actually very small. In STM images, the feedback loop is on and the tip can get closer to the sample when the DB is positive. 

Because the sample is n-type, the DB’s native charge state is negative in the absence of the STM tip \cite{Liu.2002,Piva.2005,Bellec.2009,Haider.2009,Schofield.2013}. The slightly bright ring at the edge of the halo that is more clearly seen as an increase in the apparent height profile is an indication that the DB is negative when the STM tip is a few nm from the DB \cite{Livadaru.2011,Schofield.2013}. As the tip moves closer to the DB, the emptying rate starts to overtake the filling rate ($\Gamma_{tip}>\Gamma_{Si}$) resulting in the sudden decrease in the apparent height due to the DB that becomes predominantly positive. The speckles around the dark halo which are reminiscent of the telegraph noise seen in empty state imaging \cite{Taucer.2014} further highlight that the charge state of the DB in filled states imaging is determined by the competition between tunneling rates through the DB.

\section{First-principles calculations: influence of DB-dopant distance}
The previous sections clearly show the importance of subsurface dopants influence on the DB electronic properties. In this section, we used density functional theory (DFT) to study the effect of dopant location on the electronic structure of the DB. This effect could be an important factor to account for the variation of the charge state transition peak. Other factors such as the local surface environment can be considered as part of future detailed theoretical and experimental studies. The dopant location effect have been demonstrated by Blomquist \textit{et al.} who used a self-consistent Poisson-Schr\"{o}dinger Tight Binding model of large ($>500$ atom) silicon clusters with single phosphorus dopant \cite{Blomquist.2006}. Here, using a different method and considering an arsenic dopant as used in our experiments, we obtain qualitatively similar results.  

The DFT calculations have been performed using the VASP code with projector augmented-wave (PAW) method \cite{kresse1996efficient,kresse1999ultrasoft}. The periodic super-cell approach is employed to model a $(100)-(2\times1):H$ silicon slab containing three dimer rows with six dimers per row. The slab consists of 
eighteen Si layers with an additional $1.2\,nm$ of vacuum to reduce the interaction between the periodic images. The bottom layer of slab is terminated with hydrogen atoms. GGA-PBE (Generalized Gradient Approximation-Perdew-Burke-Ernzerhof) is used for the exchange-correlation functional \cite{perdew1996generalized} with a $2\times2$ \textit{k}-mesh for Brillouin-zone sampling. $250\,eV$ has been set as the cutoff energy for plane-wave expansion of the wave function. The structures have been relaxed with a force threshold of $0.02\,eV$/\AA. In order to increase the accuracy of the calculated energy levels, density of states (DOS) have been obtained by subsequent non-selfconsistent calculations using a $4\times4$ \textit{k}-mesh. The PDOS is obtained by a quick projection scheme based on the PAW formalism as implemented in the VASP code. These calculations were performed in the absence of an STM tip.

\begin{figure*}[ht]
	\centering
	\includegraphics[width=0.6\textwidth]{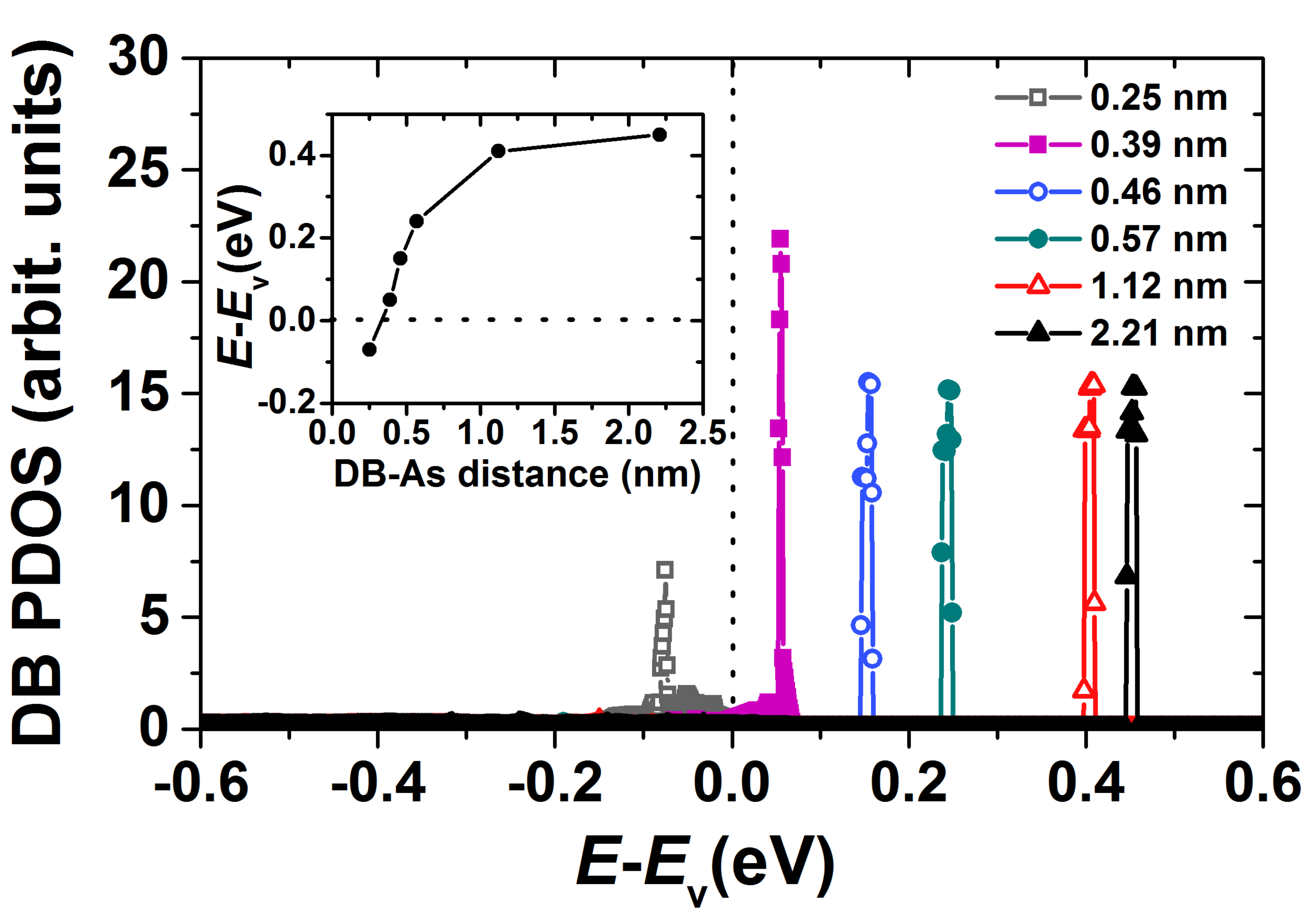}
	\caption[PDOS]{Partial density of states (PDOS) of a negatively charged DB for different 
		distances between arsenic dopant and DB. The inset shows the energy dependence 
		of the DB state on this distance.} 
	\label{Fig6}
\end{figure*}

Six slabs have been modeled in which one arsenic dopant substitutes a silicon atom at different sites below the silicon atom hosting the DB. The dopant site ranges from the DB's first nearest neighbor at $\sim0.2\,nm$ until  $\sim2\,nm$  apart from the DB atom into the bulk. The DB PDOS for each case is depicted in figure \ref{Fig6}, where their VBM are shifted 
to zero. It shows that the DB state starts from about $\sim0.4\,eV$ above the VBM for arsenic in bulk moving toward and finally merging with the valence band as the distance between DB and arsenic atom decreases. The inset of figure \ref{Fig6} shows the energy dependence of the DB state on the arsenic atom location. 

Even though these calculations show the DB negative state, we can reasonably expect that the DB-As distance will also affect the DB charge transition levels (CTL). In this perspective, a more detailed theoretical study involving calculation of DB's  CTLs would be very interesting but is beyond the scope of this article. Nonetheless, the calculations presented here can explain some experimental observations. In fact, since the subsurface dopant distribution is not uniform, the influence of the DB-As distance on the DB state can explain the variability seen in the charge state transition voltage threshold determined from I(V) spectroscopy. However, it must be noted that no quantitative comparison with the experimental data can be drawn here since these DFT results show a trend for the evolution of DB state as a function of only the DB-dopant distance. Other experimental aspects that could influence the charge state transition peak such as local variation of subsurface dopant concentration, TIBB and local surface defects are not considered in these DFT calculations

\section{Discussion of a mechanism for the DB charge state transition}
It is important to stress that the peak observed in $dI/dV$ should not be understood within the usual framework of STS, that is simply as a peak in the local density of states (LDOS). This is clear from the topographic images taken at biases near the peak position, shown in Figure \ref{Fig5}. At biases slightly more positive than the peak position, where current through the DB is very small, the topography of the surrounding silicon clearly shows band bending effects consistent with a positively charged DB which implies a tip Fermi level below the DB energy level. This rules out the usual interpretation of peaks in STS being simply related to peaks in the LDOS. The observed peak must therefore be understood in terms of non-equilibrium current through an isolated energy level.

The current through a single level can be expressed as \cite{datta2004}
\begin{equation}
I = \frac{q}{h} \int_{-\infty}^{\infty} T(E) \left[ f(E;E_{F}^{tip})- f(E;E_{F}^{Si})  \right] dE ,
\end{equation}
where $f(E;E_{F})$ represent the Fermi functions for the tip and sample, as a function of the Fermi level $E_{F}^{tip,Si}$. $T(E)$ represents the transmission function, given by
\begin{equation}
T(E) = 2 \pi D_{DB}(E) \frac{\Gamma_{tip}\Gamma_{Si}}{\Gamma_{tip} + \Gamma_{Si}} ,
\end{equation}
where $D_{DB} (E)$ represents the density of states of the DB, which tends to a delta function in the limit of no coupling, or may  experience a broadening as a result of coupling to the tip and sample states. $\Gamma_{tip}$ and $\Gamma_{Si}$ represent the coupling of the DB level to the tip and sample, in units of energy, and may be functions of the energy, $E$, as well. In the limit of weak coupling, we have $D_{DB}(E) = \delta(E-E_{DB})$ and 
\begin{equation}
I = \frac{q}{\hbar} \frac{\Gamma_{tip}\Gamma_{Si}}{\Gamma_{tip}+\Gamma_{Si}}.
\label{eqn:WeakCoupling}
\end{equation}
Note that the quantity $\frac{1}{\hbar} \Gamma_{tip}\Gamma_{Si} / (\Gamma_{tip}+\Gamma_{Si})$ represents the overall rate for a process consisting of two sequential processes characterized by the rates of transfer of electrons between the DB and the tip on the one hand, and the DB and the sample on the other, $\Gamma_{tip} / \hbar$ and $\Gamma_{Si} / \hbar$. Equation 3 presents multiple possibilities for explaining the observed peak in $dI/dV$. Two broad mechanisms are discussed hereafter:

The first mechanism which can explain the observed peak involves the DB being inelastically filled with electrons from the CB and electron donors (the process labeled (i) in Figure \ref{Fig3}-b). We refer to this model as the CB model. In this view, the step in $I(V)$ can be understood by considering the supply of electrons to the DB from the bulk side, $\Gamma_{Si}$, to be rate limiting. In the limit where $\Gamma_{Si} \ll \Gamma_{tip}$, equation \ref{eqn:WeakCoupling} can be reduced to $I=q \Gamma_{Si} / \hbar$: the current is limited by the slower rate and the DB is predominantly positive. 
As the sample voltage is made more negative and the TIBB is increased,  $\Gamma_{Si}$ rapidly increases, approaches and eventually overtakes $\Gamma_{tip}$ which leads to a step in $I(V)$. The steady-state occupation of the DB changes, explaining the observed charge state transition. Transport through the donor-depleted region at $4.5\,K$, via the CB and donor levels, is non-trivial and will not be discussed in detail here. Instead, we simply note that since the donor band does not reach to the surface, the bound electrons at near-surface donors may be well localized. Conduction through the donors and CB under the influence of the tip field could be suppressed for small biases, where the field penetration into the silicon is weak, but current to the DB could very rapidly increase as the tip field increases and penetrates further into the bulk silicon. Such behaviour could, for instance, result from the well-studied sharp onset of field-ionization of donors \cite{smit2003,debernardi2006,Teichmann.2008}, or other processes related to transport through the disordered donor network in the near-surface dopant depletion region.

A second mechanism involves a current due to resonant tunneling from the VB to the DB (labelled (ii) in Figure \ref{Fig3}(b)). This filling rate is zero until the downward surface band bending is sufficient for the DB level to become resonant with the distant VB edge. As bands bend further, this rate increases as the triangular barrier through which the electron must tunnel becomes sharp. Eventually, $\Gamma_{Si}$ could overtake $\Gamma_{tip}$ via this mechanism, which would then becomes the rate limiting process. The non-linear increase in $\Gamma_{Si}$ with respect to applied bias, along with the limit set by $\Gamma_{tip}$, leads to a step in total current given in Equation \ref{eqn:WeakCoupling}. We refer to this scheme as the the VB model and note that this mode of transport injects holes into the silicon. This mechanism would contribute to filling transition levels which are close to the VB edge, but seems unlikely to explain the observed transition from a predominantly positive to a predominantly negative DB, as described here, since the (0/-) transition level is expected to be further from the VB edge. Furthermore, this model can only explain current through the DB for sample biases where the Fermi level is below the silicon bandgap ($\lesssim -1.1\,{V}$). For some DBs, we have observed the peak in $dI/dV$ to be located near or above the VB edge, which again suggests that the dominant source of current to the DB from the bulk silicon comes from the CB and donors. 

While detailed modeling of the charge state transition mechanism is beyond the scope of this paper, it is clear that the peak in dI/dV should not be understood using standard STS concepts, but instead needs to be viewed as the result of rate-limiting processes which determine a non-equilibrium current. In either of the mechanisms described above, the shift in peak position with increasing flashing of the sample is readily understood. Also, the variation from DB to DB on a single sample is understood as being related to the randomness in dopant position. Likely the overgrowth of several nm of un-doped silicon will greatly reduce variability among DBs.

\section{Conclusion and perspectives}
Through this study, we established the importance of subsurface dopants on the electronic properties of single DBs at low temperature on highly doped samples. We showed that when a subsurface dopant depletion region is formed due to the thermal treatment of the sample, single DBs exhibit a sharp charge state transition that can be clearly identified from a characteristic I(V) spectroscopy and correlated with filled states STM images at different voltages. DFT calculations further highlight the influence of subsurface dopants on the DB state.

This work further demonstrates the pertinence of considering the DB as a single atom quantum dot and opens the door for further perspectives of DB based devices. Additional experiments and a more detailed theoretical study of single DBs will investigate exact mechanisms of transport through the DB and charge state transition.

\textbf{Acknowledgments}\\
We thank NRC, NSERC and AITF for financial support. Computational resources were provided by 
WestGrid.

\bibliography{Spectro_DB}{}

\end{document}